\documentclass[useAMS,usenatbib,letterpaper]{mn2e}
\usepackage{graphicx}

\newcommand{\lsim}{ \lower .75ex \hbox{$\sim$} \llap{\raise .27ex \hbox{$<$} } }

\newcommand{\apj}{ApJ}
\newcommand{\aj}{AJ}
\newcommand{\aap}{A\&A}
\newcommand{\nat}{Nature}
\newcommand{\mnras}{MNRAS}
\newcommand{\apjl}{ApJL}
\newcommand{\apjs}{ApJS}
\newcommand{\araa}{ARA\&A}

\title[On the nature of the short duration GRB 050906]
{On the nature of the short duration GRB 050906}

\author[A.J.~Levan et al.]{A. J.~Levan$^{1}$\thanks{email: a.j.levan@warwick.ac.uk, 
Based on observations made with ESO telescopes at the Paranal Observatory under 
programme ID 075.D-0261}, 
N.R. Tanvir$^{2}$,
P. Jakobsson$^{3,4}$,
R. Chapman$^{3}$,
J. Hjorth$^{4}$,
R.S. Priddey$^{3}$, \and
J.P.U Fynbo$^{4}$,
K. Hurley$^{5}$,
B.L. Jensen$^{4}$,
R. Johnson$^{6}$,
J. Gorosabel$^{7}$,
A.J. Castro-Tirado$^{7}$, \and
M. Jarvis$^{3}$, 
D. Watson$^{4}$,
K. Wiersema$^{8}$ \\
$^{1}${Department of Physics, University of Warwick, Coventry, CV4 7AL, UK} \\
$^{2}${Department of Physics and Astronomy, University of Leicester, Leicester, LE1 7RH, UK} \\
$^{3}${Centre for Astrophysics Research, University of Hertfordshire, College Lane, Hatfield, AL10 9AB, UK} \\
$^{4}${Dark Cosmology Centre, Niels Bohr Institute, University of Copenhagen, Juliane Maries Vej 30, DK-2100 Copenhagen, Denmark} \\
$^{5}${Oxford Astrophysics, Department of Physics, University of Oxford, Keble Road, Oxford, OX1 3RH, UK} \\
$^{6}${University of California at Berkeley, Space Sciences Laboratory, CA 94720-7450} \\
$^{7}${Instituto de Astrofisica de Andalucia, 18008 Granada, Spain} \\
$^{8}${Astronomical Instituut "Anton Pannekoek" , Kruislaan 403, 1098 SJ 
Amsterdam, NL} \\
}

\begin{document}

\date{Accepted 2007 May 08. Received 2007 May 03; in original form 2007 March 12}

\pagerange{\pageref{firstpage}--\pageref{lastpage}} \pubyear{2007}

\maketitle

\label{firstpage}

\begin{abstract}
We present deep optical and infrared observations of the short duration
GRB 050906. Although no X-ray or optical/IR afterglow was discovered 
to deep limits, the
error circle of the GRB (as derived from the {\it Swift} BAT) is unusual in
containing the relatively local starburst galaxy IC328. This makes GRB 050906
a candidate burst from a soft-gamma repeater, similar to the giant flare
from SGR 1806-20. 
The probability
of chance alignment of a given BAT position with such a galaxy is small
($\lsim 1$\%), although the size of the error circle (2.6 arcminute radius)
is such that a higher $z$ origin can't be ruled out.
Indeed, the error circle also includes a moderately rich galaxy cluster
at $z=0.43$, which is a plausible location for the burst given the
apparent preference that short GRBs have for regions of high 
mass density. 
No residual optical or infrared emission has
been observed, either in the form of an afterglow or later time
emission from any associated supernova-like event. We discuss the
constraints these limits place on the progenitor of GRB 050906 based
on the expected optical signatures from both SGRs and merging compact
object systems.

\end{abstract}


\section{Introduction}

Until recently the revolution of our knowledge of gamma-ray burst
(GRB) sources was limited almost exclusively to those with durations of
$t_{90}>$2s -- so called long bursts (see e.g 
Meszaros 2006 for a review) . The discovery of afterglows 
of the long duration bursts enabled rapid progress by allowing the
identification of redshifts (e.g. Metzger et al. 1997), star forming
host galaxies (e.g. Conselice et al. 2005; Wainwright et al. 2005;
Fruchter et al. 2006) and
ultimately unambiguous supernova signatures (eg. Hjorth et al. 2003)
-- finally linking long-duration GRBs to the
collapse of massive stars.

Afterglows of the short-duration GRBs (S-GRBs)  have still only
been discovered for relatively few bursts (e.g. Gehrels et al. 2005; Bloom et
al. 2005; Hjorth et al. 2005a, Fox et al. 2005; Berger et al. 2005;
Soderberg et al. 2006; Levan et al. 2006a; Levan \& Hjorth 2006).
Nonetheless the afterglows (e.g. Hjorth et al. 2005a; Fox et al. 2005; Burrows et al. 2006; Grupe et al. 2006; Campana et al. 2006), and the host galaxies which they select (e.g Gal-Yam et
al. 2005; Barthelmy et al. 2005; Prochaska et al. 2005; Gorosabel et al. 2006)
are apparently different, at least on the average, 
to those of long-duration GRBs.
Short GRBs seem to occur in galaxies
of all types including those with a older stellar populations, although
the statistical significance of such associations is
often fairly low (see e.g. Bloom et al. 2007; Levan et al. 2007). S-GRBs
are typically located at larger distances from their host galaxy
nuclei and are also less luminous and at a lower mean redshift
than long-duration bursts
(cf. Jakobsson et al. 2006), although at least one short-burst (GRB
060121; Levan et al. 2006a; de Ugarte Postigo et al. 2006) apparently originates from significantly
higher redshift, and may point to the existence of
a larger population
of high redshift S-GRBs.
These properties can naturally be explained as being
due to the merger of a tight binary consisting of
compact objects (neutron stars (NS) or black holes (BH)) following energy and angular
momentum dissipation via gravitational radiation (eg. Rosswog \&
Ramirez-Ruiz 2003; Rosswog, Ramirez-Ruiz \& Davies 2003;
Davies, Levan \& King 2005), although this is by no means the only
viable mechanism.

However, while the properties described above are already diverse
they may well not represent the whole S-GRB population.
The discovery of a massive flare from soft gamma-ray repeater
(SGR) 1806-20 (Hurley et al. 2005; Palmer et al. 2005) provided
evidence that some fraction of the 
large sample of S-GRBs found by the 
Burst And Transient Source
Experiment (BATSE) 
could be explained by SGR
giant-flares in galaxies out to $\sim 30-40$ Mpc, and
potentially further with {\it Swift}.
Furthermore a correlation of short bursts detected by BATSE with
galaxies within the local universe ($<100$Mpc) reveals that a fraction
(between 10-25\%) of short bursts originate from this nearby
large-scale structure (Tanvir et al. 2005).  Plausible (but broad)
luminosity functions can accommodate both a moderate fraction of
bursts in the local universe and a significant fraction at $z>0.2$
(Nakar et al. 2006) even if they are from a
single class of progenitor.  Perhaps rather more likely is that two
populations of {\it short}-bursts are being observed, those from SGR
giant flares, and those due to other events, most likely NS-NS or NS-BH
mergers.

To date it has not been possible to identify with high confidence the individual host
galaxies of short bursts which may be due to SGR giant flares, due
to the large error regions associated with BATSE bursts and the
relative dearth of smaller error boxes from eg. the interplanetary
network (IPN - Hurley et al. 2002). Here we present optical observations of the field of
{\it Swift}-discovered GRB 050906 (Krimm et al. 2005). The bright, nearby galaxy IC~328
lies within its positional error circle and makes a good case for a short
GRB associated with an SGR giant flare. 
However, as we show, a higher redshift origin can't be ruled out,
and in particular
a bright galaxy
cluster at $z=0.43$ also overlaps  the error circle and provides a
viable alternative origin for the burst.

\section{Observations}

GRB 050906 was detected with the {\it Swift} satellite (Gehrels et al. 2004) on 
2005 September 06 10:32 UT (day 6.4389).
The on-board reported location was RA = 03$^h$ 31$^m$
13$^s$, Dec = -14$^{\circ}$ 37\arcmin\ 30\arcsec, with a positional
accuracy of 3\arcmin (Krimm et al. 2005). As the burst was very faint it was not at first
clear if a real GRB had been observed. However it was pointed out
immediately that the error circle of GRB 050906 was unusual in
containing a bright, low-redshift galaxy, IC~328 (Levan \& Tanvir
2005), which is at a redshift of $z=0.031$, a distance of only $\approx 130$ Mpc 
(Assuming a standard $\Lambda CDM$ cosmology with $\Omega_M = 0.27$, $\Omega_{\Lambda}
=0.73$, $H_0 = 73$ km s$^{-1}$ Mpc$^{-1}$). 

A prompt slew was performed to the location of GRB 050906 with
observations with the X-ray telescope (XRT - Burrows et al. 2005)
beginning 79~s after the burst.  These observations failed
to locate any convincing X-ray afterglow (Pagani et al. 2005). Subsequently the BAT localisation of
GRB 050906 was refined to RA= 03$^h$ 31$^m$ 22$^s$, Dec= -14$^{\circ}$
39\arcmin\ 00\arcsec, with a 2.6 arcmin (90\%) uncertainty (Parsons et
al. 2005).  Its $t_{90}$ duration was 128 ms and the total fluence in
the 15-150 keV band was very low at $7.0 \pm 3.2 \times 10^{-9}$ ergs
cm$^{-2}$. No prompt observations were obtained with the UV/Optical
telescope (UVOT, Roming et al. 2005) since it
was in safe mode at the time of the burst.  Full details of the {\it
 Swift} observations of GRB 050906 are described in a separate paper
(Hurley et al. in prep), which concludes that the burst would be surprisingly
soft if from a SGR giant flare.  However, since the properties of
giant flares are rather poorly understood, it remains important to
look at the other evidence for and against such an explanation
in this particular case.

We first observed the error circle of GRB 050906 using the Wide Field
Camera (WFCAM) on the United Kingdom Infrared Telescope (UKIRT)
beginning at Sept 6.51(UT), 1.7 hours after the burst (a complete
log is shown in Table 1). A
K-band integration of 40 minutes was made. A second, similar exposure was obtained
the following night ($\sim$ 26 hours after the burst). Deep optical
observations were acquired at the ESO Very Large Telescope (VLT),
covering the VRI bands ($\sim$1800s in each band).  Three epochs of
observations were obtained using the FORS1 and FORS2 imagers at 
21 and 140 hours and 18 days after the burst. We also obtained deep, late
time IR observations using VLT/ISAAC in J,H and K. Optical
observations were processed through IRAF in the standard fashion. The
UKIRT/WFCAM observations were reduced via the ORAC-DR pipeline
(Cavanagh et al. 2003) and the VLT/ISAAC observations were processed
with {\it eclipse} (Devillard 1997).

Cameron \& Frail (2005) identified a single radio source within the
initial burst error circle (although outside the refined BAT source
location), the position of this source is RA = 03$^h$ 31$^m$ 11.8$^s$,
Dec = -14$^{\circ}$ 37\arcmin 18.1\arcsec.  At this location our
images reveal a very red point source ($R-K = 5.6$). However, it did not
exhibit any variability in the optical or IR and it is likely to be a
background galaxy.

Although the prompt XRT X-ray observations failed to locate any strong candidate
for the afterglow of the burst, an inspection of the images did produce
a possible faint counterpart (Fox et al. 2005; Butler 2006).  However,
its association with GRB 050906, and even its reality remain highly uncertain.
The refined location 
of this candidate is  RA = 03$^h$ 31$^m$ 15.28$^s$,
Dec = -14$^{\circ}$ 36\arcmin 13.1\arcsec. We find no evidence
for any variable point sources within this region, nor for
any particular overdensity of galaxies within the large (15.7 \arcsec
radius) localisation.

A detailed inspection of the entire error region within our 
observations revealed no evidence for new sources
by comparison to archival surveys, or between our own images taken at
different epochs.  We estimate the limiting magnitude of each
individual frame by examining the signal-to-noise ratios for the
photometry of many point sources: the resulting limiting
magnitudes are shown in Table 1. In order to search for a variable
afterglow (which may be placed on top of a relatively bright host
galaxy) we performed PSF-matched image subtractions of the different
epochs of imaging using the code of Alard \& Lupton (1998). Each epoch
of observations was subtracted from each other epoch obtained in the
same filter.  These subtractions yielded very clean residual images in
the cases of using the same telescope and instrument, although larger
residuals where observations had to be matched from different
telescopes. To estimate the limiting magnitude of any variable sources
we seeded each image with a number of false stars
(which were added with the appropriate PSF for the image in which they
are seeded) and then repeated the subtractions.  The magnitude of
sources which can be recovered as residuals at $> 5 \sigma$ in the
resulting difference images are also shown in Table 1.

As the error circle of GRB 050906 contains the bright nearby galaxy IC
328 (and the companion galaxy IC 327 is just outside the error circle) we
separately estimate the limiting magnitude for any variable source
within the galaxy. These limits are lower than for the field in
general since the bright cores of each galaxy leave large residuals in
the subtracted images. Any source within 5\arcsec of the
centre of either galaxy would have to be $R<20$ in order to be
detected clearly in our residual images. Although this is relatively
bright it should be noted that even a moderately faint supernova in IC~328 
(e.g. one with $M_V \sim -17$ at maximum), if unextinguished, would reach a
peak about 2.5 magnitudes brighter than this.

\begin{figure}
\centerline{
\resizebox{8truecm}{!}{\includegraphics[angle=0]{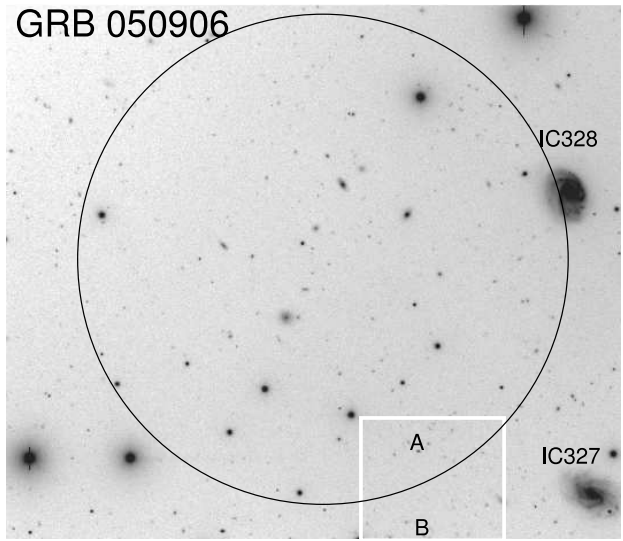}}}
\caption[]{Finding chart for GRB 050906. North is up and east 
to the left. The GRB error circle of 5.2\arcmin\ diameter is shown. 
As can be seen the bright, low-redshift galaxy IC~328 lies on the edge of this 
error circle, which would be an unlikely chance coincidence if it
is not associated with the GRB.  
However, there are many other more distant galaxies clearly seen
in the field, including a galaxy cluster which, as discussed in the
text, is also a viable location of the GRB. Figure 2 shows an enlarged region 
about this cluster 
and the area shown is marked by the white box.}
\label{fig2} 
\end{figure}

\begin{figure}
\centerline{
\resizebox{8truecm}{!}{\includegraphics[angle=0]{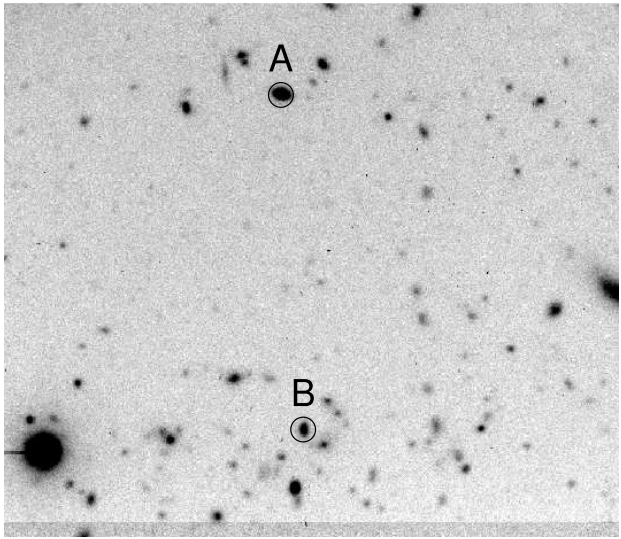}}}
\caption[]{A $z=0.43$ cluster lying within the error circle of GRB 050906. 
The two galaxies marked A \& B have spectroscopic redshifts, while the 
other galaxies exhibit similar colours and
are likely to lie within the same cluster.}
\label{fig2} 
\end{figure}

In addition to IC~328 and IC~327, visual inspection of our FORS images
of GRB~050906 revealed many more distant galaxies, including an
overdensity  of galaxies in the south west of the error circle. 
Although
this clustering extended both beyond the error circle and further
beyond the field of view of our FORS observations, we estimate that
the greatest concentration lies at roughly RA = 03$^h$ 31$^m$ 17$^s$, Dec =
-14$^{\circ}$ 41\arcmin 25\arcsec,
although the distribution of galaxies is not
uniform and exhibits at least two (possibly more) regions of overdensity
(see Figure 2).

We obtained spectroscopy of two galaxies from this concentration using
Gemini South and GMOS on 2006 January 26, with the G300V grating. The
galaxies in question are marked in Figures 1 \& 2.  Spectra were
reduced in the standard fashion using the specific GMOS scripts
within IRAF. Inspection of these
spectra reveals strong absorption features which we attribute to Ca H
\& K, and H$\delta$ at a redshift of $z=0.43$.

\begin{table*}
\begin{center}
\begin{tabular}{llllll}
\hline
Date & $\Delta t_b$ (days) &  exp. time (s) &  Instrument/Filter & Limit (Frame) & Limit (sub) \\
\hline
2005 Sept 6.502		& 0.06 	& 2160 & WFCAM/K & 20.2  & 19.5 (WFCAM)\\
2005 Sept 7.333		& 0.89	& 1800&  FORS2/R &  26.6 & 25.9 (FORS2)\\
2005 Sept 7.345		& 0.91 & 1800	& FORS2/V  & 27.3  & 26.6 (FORS2)\\
2005 Sept 7.3670		& 0.93 & 1920	& FORS2/I    & 25.2 & 24.5 (FORS2)\\
2005 Sept 7.510		& 1.69  & 3240	& WFCAM/K & 20.4  & 19.5 (WFCAM)\\
2005 Sept 12.333    		& 5.89 & 1800 	& FORS2/R	& 26.4 & 25.9 (FORS2)\\
2005 Sept 12.344      	& 5.90 & 1800	& FORS2/V	& 27.1 & 26.6 (FORS2)\\
2005 Sept 12.357		& 5.92 & 1920	& FORS2/I	& 25.0 & 26.6 (FORS2)\\
2005  Sept 25.272		& 18.83 & 1800	& FORS1/R	& 26.2 & 25.2 (FORS2)\\
2005 Sept 25.280		& 18.84 &1800	& FORS1/V	& 26.5 & 25.7 (FORS2)\\
2005 Sept 25.293		& 18.85 & 	1800 & FORS1/I	& 24.9 & 24.2 (FORS2)\\
2005 Sept 29.294 		& 22.86 & 320	& ISAAC/K &	22.1        & 19.5 (WFCAM)\\
2005 Sept 29.340 		& 22.90 & 280	& ISAAC/J	  &	23.1& - \\
2005 Sept 29.346 		& 22.91 & 208 	& ISAAC/H &	22.2		& - \\
\hline
\label{tab:obslog}
\end{tabular}
\end{center}
\caption{Optical/Infrared observations of GRB 050906 
obtained at UKIRT
and the VLT. The date of the start of the observations is shown as is the time since burst
and the individual frame limit. The limit for the subtractions was determined 
via the use of artificial stars, the name in parenthesis is the instrument 
used to perform the subtraction which yielded this limit. }
\label{default}
\end{table*}%

\begin{table*}
\begin{center}
\begin{tabular}{llll}
\hline
Filter & IC~328 & IC~327 & references \\
\hline
U & 				14.3 & - & Coziol et al. 1994 \\
B & 				14.9	&                 15.15 $\pm$ 0.12	& Coziol et al. 1994\\
V &				14.28 $\pm $ 0.02	& 14.93 $\pm$ 0.02	& This work \\
R &				13.60 $\pm$ 0.02 	& 14.45 $\pm$ 0.02	& This work \\	
I  &				13.05 $\pm$ 0.02	& 13.95 $\pm$ 0.02	& This work \\
J & 				12.48 $\pm$ 0.02	& 12.90 $\pm$ 0.05 & 2MASS \\ 
H & 				11.64 $\pm$  0.05	& 12.37 $\pm$ 0.08 &2MASS \\
K & 				11.35 $\pm$ 0.06 	& 12.13 $\pm$ 0.10 & 2MASS \\
IRAS 12 $\mu$ m & 	$(9.36 \pm 2.15) \times 10^{-2}$ Jy	&  $<9.32 \times 10^{-2}$ Jy &Moshir et al. 1990 \\
IRAS 25 $\mu$ m &	$(8.26 \pm 2.25) \times 10^{-2}$ Jy	&  $<7.70 \times 10^{-2}$ Jy &Moshir et al. 1990 \\
IRAS 60 $\mu$ m & 	$0.80 \pm 0.05$ Jy	& 0.21 $\pm$ 0.04 Jy &  Moshir et al. 1990 \\
IRAS 100 $\mu$ m & $2.23 \pm 0.29$ Jy	& $< 2.13$ Jy & Moshir et al. 1990 \\
1.4 GHz			& -				& $2.7 \pm 0.5$ mJy & Condon et al. 1998 \\
\hline
\end{tabular}
\end{center}
\caption{Optical/Infrared observations of IC~328 and IC~327 obtained from
the literature as cited and via our VLT observations.  The optical/nIR photometry has been
corrected for foreground extinction following Schlegel et al. (1998).}
\label{tab:photdata}
\end{table*}%

\section{The properties of IC~328}

IC~328 is a bright K=11.4 galaxy and its colours,
and somewhat disturbed morphology, are consistent with an actively
star forming, late-type galaxy. 
Its observed B-band magnitude (B$_0$=14.9,
$M_B$=-20.7) is approximately L* and its R-K colour of $\approx 2.3$ is very
blue, also indicative of ongoing star formation.  IC~328 was detected
by IRAS in all four bands (10,25,60,100 $\mu m$).  Converting from the
observed 60 $\mu m$ flux to a star formation rate assuming a
relation of $SFR = 5.5 * (L_{60} / 5.1 \times 10^{30} {\rm ergs/s/Hz})$
(Kennicutt 1998) results in a star formation rate for IC~328 of
$\sim$17 M$_{\odot}$ yr$^{-1}$.  Figure 3 shows the SED of IC~328 
(see also Table~\ref{tab:photdata}) overlayed 
with several comparison spectra (standard Sc, M51
and M82).  An alternative means of estimating the star formation rates
is to scale these template spectra such that they provide a reasonable
fit to the observed spectral energy distribution of IC~328. Doing 
this with M51 yields a star formation rate of $\sim 3$ M$_{\odot}$ yr$^{-1}$,
significantly lower than via the 60 $\mu m$ flux, but consistent
with the idea that much of the star formation in IC~328 is dust-obscured,
leading to the high fIR fluxes. 

SGRs are commonly thought to be formed via the core-collapse of
massive stars and thus would trace the star formation rate of a given
galaxy, although they may also be produced via accretion induced
collapse (AIC) of merging white dwarfs (Usov 1992; Levan et al. 2006b)
and SGRs formed via this channel should trace the stellar mass
density. The IR luminosity of IC~328 implies that it is moderately
massive: using the stellar mass estimation scheme of Mannucci et al. (2005)
yields $M \sim 10^{11}$ M$_{\odot}$.  Thus IC~328 appears in
terms of mass to be similar to the Milky Way, and in terms of star
formation rather more active than the MW by factors of several, and
hence can be expected to harbour at least a similar number of SGRs.

If originating from IC~328 the isotropic equivalent energy of GRB 050906 would be
$E_{iso} \sim 1.5 \times 10^{46}$ ergs in the 15-150 keV range.
This compares to a total energy release ($>30$ keV) of $E_{iso} \sim 4 \times
10^{46}$ ergs for the giant flare from SGR 1806-20 (Hurley et al. 2005).

\begin{figure}
\centerline{
\resizebox{8truecm}{!}{\includegraphics[angle=0]{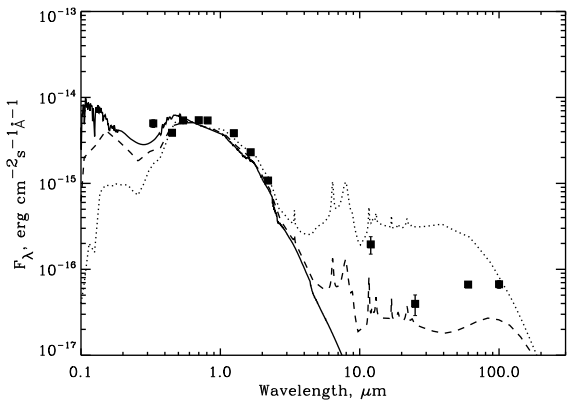}}}
\caption[]{The spectral energy distribution of IC~328. Overlayed with template 
spectra for an Sc galaxy (solid line), M51 (dashed line) and M82 (dotted line). 
The spectra have been normalised to the R-band flux. Extrapolating from the M51
template we thus obtain a star formation rate of 3 M$_{\odot}$ yr$^{-1}$,
somewhat lower than that that inferred from the 60 $\mu m$ flux.}
\label{fig2} 
\end{figure}

Galaxies such as IC~328 are rare in GRB error circles only a few
arcminutes in diameter. To estimate the probability of a chance
alignment we simulated a large set of GRB positions placed randomly on
the sky and subsequently searched for galaxies within 3 arcminutes of
these positions (our galaxy catalogue was the complete IRAS PSCz
catalogue, which contains IC~328 and is thus a good catalogue to
search for similar galaxies). In a total of 50,000 random bursts only
135 matches were found, allowing for the 15\% avoidance of the
Galactic plane in the PSCz survey this implies a probability of
selecting IR-bright galaxies such as IC~328 of only 0.003.  There are,
of course, alternative ways this probability analysis could have been
performed, for instance using optically selected galaxy catalogues.
However we feel that our approach is suitably conservative (eg. we
might have cut the PSCz sample to galaxies as bright as IC~328 or
brighter), and therefore gives a useful indication of the low
likelihood of a chance coincidence.

We emphasise that this probability should not simply be regarded as an
{a posteriori} calculation. Tanvir et al. (2005) had already predicted
that a non-negligible proportion of short-duration bursts should be
associated with low-redshift galaxies, so it is statistically
reasonable, therefore, to specifically test the null-hypothesis that
there is no such association.  At the time of writing roughly a dozen short
bursts have been observed by {\it Swift}, and a similar number
previously well-localised
by IPN and HETE-2/BeppoSAX (although those detected by the
IPN have a significantly different selection function so comparing
their properties with {\it HETE-2/SXC} and {\it Swift} bursts is
non-trivial). In any event, based on our analysis above, we can say
with some confidence that the probability of a {\em chance} occurence
of such a nearby and bright galaxy as IC~328 in {\em one or more} of
the {\it HETE-2}/SXC and {\it Swift}/BAT burst error regions is less than 10\%.

Further, although none of the other short bursts detected by Swift have        
plausible local hosts, the IPN has delivered locations for two S-GRBs that      
may originate in very local galaxies.
Firstly GRB 051103 has an 
error box which overlaps the outskirts of
both M81 and M82 (Fredericks et al. 2006; Ofek et al. 2006), 
while the recent GRB 070201 has an error box which intersects the spiral arms of
M31 (Golenetskii et al. 2007; Hurley et al. 2007).
These locations lend support to the results of Tanvir et al. (2005)
that a fraction of short bursts should originate in the local universe, and
give further credence to the suggestion that IC~328 is the host galaxy
of GRB~050906. 

\begin{figure}
\centerline{
\resizebox{8truecm}{!}{\includegraphics[angle=0]{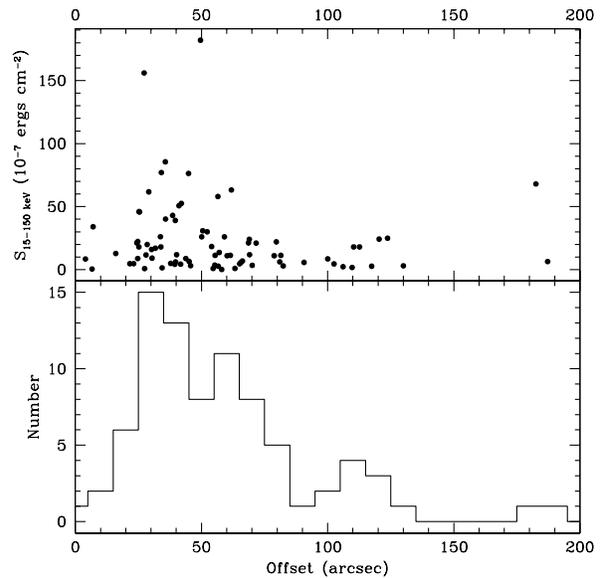}}}
\caption[]{The offset distribution between the refined BAT localisations
and the positions of the X-ray afterglows of GRBs detected by {\it Swift}.
GRBs have been plotted for the period of one full year of {\it Swift} 
operations, from 1 April 2005 - 30 March 2006 (94 bursts with
X-ray afterglow locations). As can be seen the majority
of {\it Swift} bursts lie relatively close to the refined BAT position.
Formally 67\% lie within 62\arcsec\ and 90\% lie within 109\arcsec.
The probability of locating an afterglow greater than 160\arcsec\ from the
burst position is $\sim$2\%, and is dominated by the two outliers, GRB 060109 and
GRB 060218.  There is perhaps a slight trend towards larger offsets for
fainter bursts on the average.}
\label{fig:offsets} 
\end{figure}

However, it is also important to ask what is the probability of
finding IC~328 at the position we do in the error circle if it is truly
associated with the burst.  Formally, only a small fraction of IC~328
lies within the refined BAT error circle, which, although nominally 90\%
confidence, are typically conservative (Fenimore, private
communication), although this may not be the case
for the very short and faint GRB 050906. To gauge the number of bursts which we might expect to
lie $>$2.5\arcmin\ from the BAT localisation we have plotted in
Figure~\ref{fig:offsets} the offset distribution between XRT and BAT
positions for all bursts exhibiting afterglows
to the XRT in the first year of full {\it Swift} operations.
This shows that, in fact,
90\% of
the bursts occur within 110\arcsec of the BAT localisation, and thus,
even faint bursts like GRB 050906 should rarely ($\sim2$\% of bursts)
be at the radial separation of
IC~328 from the centre of the BAT error circle. Hence, while the
association of GRB 050906 with IC~328 remains plausible, and would have
been identified as the host galaxy by various approaches (e.g. that of
Gal-Yam et al. 2006), the location at the edge of the error circle
does somewhat weaken the case for an association.

Of course, were GRB 050906 due to a low-luminosity NS-NS merger event
within IC328 (or even IC~327) then it may be expected to be at a large
distance from its parent galaxy either because it took place in the
halo (eg. a globular cluster), or having been ejected outside the main
body of the galaxy by a natal supernova kick.  A kick of 30~kpc
(e.g. similar to that inferred for GRB 050509B) would have led to an
offset of $\sim$ 1\arcmin, and could place GRB 050906 relatively
closer to the BAT localisation. Additionally the error circle contains a number 
of fainter (but still moderately bright) galaxies whose colours and physical 
sizes could well associate them with a group containing IC~328 and IC~327. 
Several of these galaxies lie relatively close to the centre of the BAT error
box.

\section{Other Galaxies within the Error Circle}
In addition to IC328 and the possibly associated galaxies described above,
there are many more fainter galaxies which probably lie
at a range of higher redshifts.  The most notable structure is a
relatively rich galaxy cluster which overlaps the southern part of the
error circle, which, as discussed in section 2, is at $z=0.43$.
Some, previously well-localised short bursts have been found to lie in
regions of high mass density (e.g. in elliptical 
galaxies or cluster environments - Gehrels
et al. 2005; Pedersen et al. 2005; Bloom et al. 2006),
although the faint, and likely high redshift, host galaxies to GRB 060121 (Levan et al. 2006a; 
de Ugarte Postigo et al. 2006) and
GRB 060313 (Hjorth et al. in prep) indicate this is not 
necessarily the case and that a significant fraction
of SGRBs may originate at higher redshift (Berger et al. 2006a).
Thus, it is certainly plausible
that GRB 050906 originated in this cluster.  Its duration of 128 ms is
comparable to that of GRB 050509B and, at $z=0.43$, its inferred
isotropic energy release would be $E_{iso} = (3.2 \pm 1.4) \times
10^{48}$ ergs, would also place it along with GRB 050509B
($t_{90}$ = 40 ms , $E_{iso} = 1 \times 10^{48}$ ergs; Gehrels et al. 
2005) as the
intrinsically faintest of the cosmological short GRBs seen to-date.

\section{Implications for Progenitor Models}
The leading contenders for the progenitors of short-duration GRBs are
those which are the result of compact
binary mergers (NS-NS, NS-BH and possibly WD-BH -- see e.g. Lee \& Ramirez-Ruiz for
a review) and those
resulting from giant flares from soft-gamma-repeaters. Observations
of S-GRBs detected by {\it Swift} provide some support for the former model, while
the giant flare from SGR 1806-20 (Hurley et al. 2005; Palmer et al. 2005)
provided renewed impetus to investigate
SGRs as candidate progenitors. In particular these may be responsible
for the fraction of S-GRBs in the local universe reported by Tanvir et al. (2005),
and for the two subsequently detected IPN bursts which may have
originated from M81/82 and M31.
The most intense
spike of the SGR 1806-20 event would have been detected by BATSE as a short-hard
gamma-ray burst had it occured out to about 30-40 Mpc
(Hurley et al. 2005; Palmer et al. 2005), and it 
would be positively surprising if some proportion of BATSE S-GRBs were not
due to such events occurring in nearby galaxies.  Most estimates of the
volume average star formation rate in the local universe put it at about
0.02 M$_{\odot}$~yr$^{-1}$~Mpc$^{-3}$ (eg. Iglesias-Paramo et al. 2006).  So in a
sphere of radius 100 Mpc we would expect to find a total rate of
star formation roughly 20000 times the current rate in the Milky Way.
SGRs are thought to be young (and short-lived), highly-magnetised neutron
stars, and so their number within a galaxy may reflect its
star-formation rate (but see Levan et al 2006b for a possible route to
creating magnetars in old stellar populations via WD-WD
mergers).  Thus, even if an SGR 1806-20-like event were only
to occur in the MW on average once every two millenia,  $\sim$10 per year
should occur within this volume.  
Of course, the SGR 1806-20 flare itself might not have been quite
luminous enough to have been detected to 100 Mpc, but equally, it is
unlikely that this event was at the very peak of the luminosity function,
and the rate of lower luminosity flares may be greater.

We must bear in mind that the observed afterglows of many
short-duration GRBs have been relatively faint in comparison to
typical long bursts (Hjorth et al. 2005a; Fox et al. 2005; Berger et
al. 2005; Soderberg et al. 2006; Levan et al. 2006). 
Although some are relatively bright, especially in X-rays
(e.g. 050724 Barthelmy et al. 2005, Campana et al. 2005;
GRB 051121 Burrows et al. 2006 and GRB 060313 Roming et al. 2006),
a fraction are very faint or undetected both in the optical
 (Hjorth et al. 2005b; Bloom et
al. 2006; Castro-Tirado et al. 2006) or X-ray
(e.g. Mineo et al. 2005; Page et al. 2006; La Parola et al. 2006). 
Similarly, SGRs appear to produce little optical/IR
emission during their giant flares, although constraints are not
strong.  This is partly due to the Galactic SGRs being in the plane of
the Milky Way and hence along dusty lines of sight, which would be
less of an issue for an observer oriented more face-on to the Galactic
plane.

SGR giant flares may produce optical emission and indeed mini-fireball
models can accurately represent the radio and X-ray emission following
the giant flare of SGR 1806-20. Being
close to the galactic plane provides a considerable challenge to
optical observations, although a candidate faint counterpart to SGR 1806-20 has been
identified from near-IR K-band observations; Kosugi et al. 2005,
Israel et al. 2005.  However the extrapolation of the expected SGR
X-ray flux into the optical waveband and the subsequent extrapolation
out to the distance of IC~328 falls below the detection limits of our
observations, especially given (i) SGRs are likely to be located close
to the nucleus of the galaxy where our limits are least constraining,
and (ii) we infer a high proportion of star formation in IC~328 is
dust obscured.

The SGR scenario is not the only possibility for this burst, since it
is plausible that another progenitor system created the GRB either in
IC 328 or a more distant galaxy. Indeed, the error
box of GRB 050906 also appears to contain a high redshift cluster, an
environment in which several short-duration GRBs have been found
(e.g. Pedersen et al. 2005; Gal-Yam et al. 2005; Berger et al. 2007).  We thus cannot rule
out a burst originating from a galaxy associated with this cluster and
consider below the implications for short bursts located in either
IC~328 or the higher-z cluster.

The principle alternative model is that of NS-NS mergers.
These might also be expected to result in bright though
relatively short lived optical emission due to the production of heavy
elements during the mergers (e.g. Li \& Paczynski 1998) -- so called
mini-SN or macro-novae (MN). These
transients can reach absolute magnitudes comparable to those of SNe,
although typically last for a much shorter duration, peaking only a
day or so after the merger (although the precise behaviour depends on
the nuclear yields in the merger itself which are only poorly
understood; but see Rosswog et al. 2000). In Figure 5 we show the
limits on any residual emission within the GRB 050906 error circle. We
also overplot the SN Ic supernova SN 2002ap at the 
distance of IC 328  ($z=0.03$).   

We do not overplot the predicted magnitudes for any MN emission
since the behaviour of
such transients is only known from theory as none have been directly
observed. However, canonical parameters thought to be associated
with NS-NS mergers (e.g. Kulkarni 2005) would predict that they would
reach peak fluxes of $\sim 0.1 \mu$Jy at $z=0.2$, or several 
$\mu$Jy ($R \sim 22$) when extrapolated to the distance of IC328.
The models also predict that they will reach this maximum on a timescale
of hours to days past the explosion, and can therefore be 
searched for in our deep optical imaging 1, 6 and 19 days post burst.
Although previous S-GRBs have not shown any sign of MN emission
(e.g. Hjorth et al. 2005a,b; Fox et al. 2005; Bloom et al. 2006),
these bursts lay at distances more than an order of magnitude
greater than IC~328, and any SN-like event occurring within them would
thus need to be significantly ($\sim 5$ magnitudes)
brighter.  Furthermore as NS-NS systems have long lifetimes and
significant natal kicks it is unlikely that a NS-NS system would be
buried within the disk of IC~328.  Therefore, while the properties
of associated MN events remain highly uncertain, given the 
current predications of their brightness it would be somewhat unexpected
that our deep observations would not uncover any indication of them should
GRB~050906 originate at the distance of IC~328. 
This lack of emission can be
remedied with NS-NS mergers if (i) the NS-NS merger expelled very
little mass or radioactive material or (ii) the true distance to GRB
050906 is significantly beyond IC~328 (e.g. in the cluster at
$z=0.43$).

\begin{figure}
\centerline{
\resizebox{8truecm}{!}{\includegraphics[angle=0]{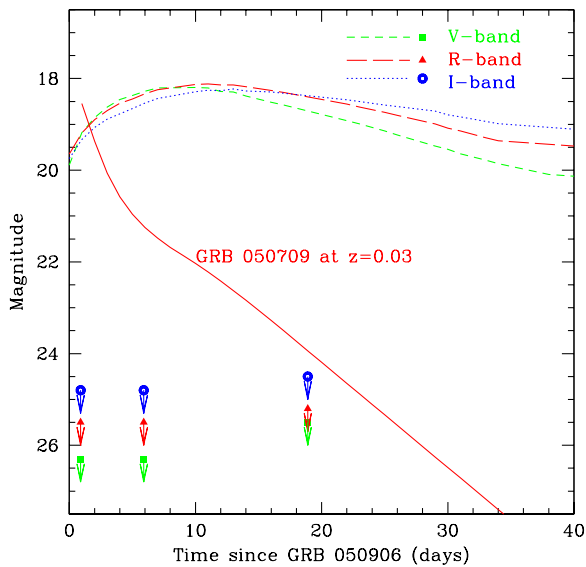}}}
\caption[]{Supernova limits in the field 
of GRB 050906 assuming that it lies at $z=0.03$. The light curves shown
are those of SN~2002ap (Foley et al. 2002) - which was a SN Ic with
a peak $M_V$=-17. Also shown is the extrapolated lightcurve of GRB 050709
as it would appear at $z=0.03$. As can be seen, the observed limits lie
significantly below these extrapolations.}
\label{fig2} 
\end{figure}

\section{Conclusions}

We have presented deep optical and infrared observations of GRB
050906, the fourth short GRB to be localised by {\it Swift} to a few
arcmin error. Although these observations fail to locate any optical
or IR afterglow (or associated SN event) to deep limits, they do
provide information on the possible environment of the burst.
Previous short bursts have been shown to be associated with a range of
host galaxy types, including those with older
stellar populations, over a wide span in redshift (eg. Berger et al. 2006;
Nakar et al. 2006).  There are indeed many distant galaxies within
the GRB 050906 positional error circle, including parts of a bright
galaxy cluster at $z=0.43$.
It is quite plausible that
GRB 050906 originates in a galaxy associated with that cluster
and is a short burst comparable to those which have already
been localised by {\it Swift}.

However, GRB 050906 is unusual in that it also
contains within its error circle a luminous local galaxy, IC~328, with
a high star formation rate.  The likelihood of such a galaxy appearing
by chance in a {\it Swift}/BAT error circle is less than 1\%.  In
addition, this galaxy is likely to host a large number of SGRs since
it is both massive and actively star forming, and thus GRB 050906 may
be the first example of a well-localised short GRB due to an SGR giant flare.
The inferred isotropic energy release at the distance of IC~328
is very comparable to that
of the initial spike in the recent SGR 1806-20 giant flare,
although GRB 050906 has a distinctly softer gamma-ray spectrum
(Hurley et al. in prep). 
The location of IC~328  at the edge of
the BAT error circle weakens but does not rule out the
case for such an association.  
Indeed, we  note that giant flares like that of SGR 1806-20 would
have to be remarkably rare events in order for them not to be present
in reasonable numbers in 
the BATSE short burst catalog, and therefore further examples are to
be expected during the lifetime of {\it Swift}.

\section*{Acknowledgements}
We thank the referee D. Grupe for constructive comments on
the manuscript. We appreciate useful discussion with Enrico Ramirez-Ruiz.
AJL \& NRT thank PPARC for postdoctoral and
senior fellowship awards. PJ acknowledges support by a Marie Curie Intra-European
Fellowship within the 6th European Community Framework
Program under contract number MEIF-CT-2006-042001.
The Dark
Cosmology Centre is funded by the DNRF. RC thanks the University of
Hertfordshire for a studentship. The research           
activities of AJC-T and JG are supported by the Spanish Ministry of 
Science     
and Education through projects AYA2004-01515 and ESP2005-07714-C03-03.

\end{document}